\documentstyle[epsf,12pt]{article}
\begin{document}

\catcode`@=11
\long\def\@caption#1[#2]#3{\par\addcontentsline{\csname
  ext@#1\endcsname}{#1}{\protect\numberline{\csname
  the#1\endcsname}{\ignorespaces #2}}\begingroup
    \small
    \@parboxrestore
    \@makecaption{\csname fnum@#1\endcsname}{\ignorespaces #3}\par
  \endgroup}
\catcode`@=12
\newcommand{\newc}{\newcommand}
\newc{\gsim}{\lower.7ex\hbox{$\;\stackrel{\textstyle>}{\sim}\;$}}
\newc{\lsim}{\lower.7ex\hbox{$\;\stackrel{\textstyle<}{\sim}\;$}}
\newc{\gev}{\,{\rm GeV}}
\newc{\mev}{\,{\rm MeV}}
\newc{\ev}{\,{\rm eV}}
\newc{\kev}{\,{\rm keV}}
\newc{\tev}{\,{\rm TeV}}
\newc{\mz}{m_Z}
\newc{\mpl}{M_{Pl}}
\newc{\chifc}{\chi_{{}_{\!F\!C}}}
\newc\order{{\cal O}}
\newc\CO{\order}
\newc\CL{{\cal L}}
\newc\CY{{\cal Y}}
\newc\CH{{\cal H}}
\newc\CM{{\cal M}}
\newc\CF{{\cal F}}
\newc\CD{{\cal D}}
\newc\CN{{\cal N}}
\newc{\eps}{\epsilon}
\newc{\re}{\mbox{Re}\,}
\newc{\im}{\mbox{Im}\,}
\newc{\invpb}{\,\mbox{pb}^{-1}}
\newc{\invfb}{\,\mbox{fb}^{-1}}
\newc{\yddiag}{{\bf D}}
\newc{\yddiagd}{{\bf D^\dagger}}
\newc{\yudiag}{{\bf U}}
\newc{\yudiagd}{{\bf U^\dagger}}
\newc{\yd}{{\bf Y_D}}
\newc{\ydd}{{\bf Y_D^\dagger}}
\newc{\yu}{{\bf Y_U}}
\newc{\yud}{{\bf Y_U^\dagger}}
\newc{\ckm}{{\bf V}}
\newc{\ckmd}{{\bf V^\dagger}}
\newc{\ckmz}{{\bf V^0}}
\newc{\ckmzd}{{\bf V^{0\dagger}}}
\newc{\X}{{\bf X}}
\newc{\bbbar}{B^0-\bar B^0}
\def\bra#1{\left\langle #1 \right|}
\def\ket#1{\left| #1 \right\rangle}
\newc{\sgn}{\mbox{sgn}\,}
\newc{\m}{{\bf m}}
\newc{\msusy}{M_{\rm SUSY}}
\newc{\munif}{M_{\rm unif}}
\newc{\slepton}{{\tilde\ell}}
\newc{\Slepton}{{\tilde L}}
\newc{\sneutrino}{{\tilde\nu}}
\newc{\selectron}{{\tilde e}}
\newc{\stau}{{\tilde\tau}}
%
%
\def\NPB#1#2#3{Nucl. Phys. {\bf B#1} (19#2) #3}
\def\PLB#1#2#3{Phys. Lett. {\bf B#1} (19#2) #3}
\def\PLBold#1#2#3{Phys. Lett. {\bf#1B} (19#2) #3}
\def\PRD#1#2#3{Phys. Rev. {\bf D#1} (19#2) #3}
\def\PRL#1#2#3{Phys. Rev. Lett. {\bf#1} (19#2) #3}
\def\PRT#1#2#3{Phys. Rep. {\bf#1} (19#2) #3}
\def\ARAA#1#2#3{Ann. Rev. Astron. Astrophys. {\bf#1} (19#2) #3}
\def\ARNP#1#2#3{Ann. Rev. Nucl. Part. Sci. {\bf#1} (19#2) #3}
\def\MPL#1#2#3{Mod. Phys. Lett. {\bf #1} (19#2) #3}
\def\ZPC#1#2#3{Zeit. f\"ur Physik {\bf C#1} (19#2) #3}
\def\APJ#1#2#3{Ap. J. {\bf #1} (19#2) #3}
\def\AP#1#2#3{{Ann. Phys. } {\bf #1} (19#2) #3}
\def\RMP#1#2#3{{Rev. Mod. Phys. } {\bf #1} (19#2) #3}
\def\CMP#1#2#3{{Comm. Math. Phys. } {\bf #1} (19#2) #3}
\relax
%
%
%
\def\beq{\begin{equation}}
\def\eeq{\end{equation}}
\def\bea{\begin{eqnarray}}
\def\eea{\end{eqnarray}}
%
%
%
\newc{\ie}{{\it i.e.}}          \newc{\etal}{{\it et al.}}
\newc{\eg}{{\it e.g.}}          \newc{\etc}{{\it etc.}}
\newc{\cf}{{\it c.f.}}
\def\smuon{{\tilde\mu}}
\def\neut{{\tilde N}}
\def\char{{\tilde C}}
\def\bino{{\tilde B}}
\def\wino{{\tilde W}}
\def\higgsino{{\tilde H}}
\def\sneut{{\tilde\nu}}
%
%
%
%
\def\slash#1{\rlap{$#1$}/} 
\def\Dsl{\,\raise.15ex\hbox{/}\mkern-13.5mu D} 
\def\delsl{\raise.15ex\hbox{/}\kern-.57em\partial}
\def\Ksl{\hbox{/\kern-.6000em\rm K}}
\def\Asl{\hbox{/\kern-.6500em \rm A}}
\def\Qsl{\hbox{/\kern-.6000em\rm Q}}
\def\gradsl{\hbox{/\kern-.6500em$\nabla$}}
%
%
%
\def\bar#1{\overline{#1}}
\def\vev#1{\left\langle #1 \right\rangle}
%

\begin{titlepage} 
\begin{flushright}
CAPCyR-BUAP/04-01\\
FCFM-BUAP-HEP/04-01\\
IFUAP-HEP/04-01\\
January 2004\\
\end{flushright}
\vskip 2cm
\begin{center}
{\large\bf Mass matrix {\it Ansatz} and lepton flavor violation in the THDM-III}
\vskip 1cm
{\normalsize\bf
 J.L. D\'{\i}az-Cruz$^{(a,b)}$,  R. Noriega-Papaqui$^{(c)}$ 
and  A. Rosado$^{(a,c)}$ \\
\vskip 0.5cm
$^{(a)}${\it Cuerpo Acad\'emico de Part\'{\i}culas, Campos y Relatividad\\
de la Benem\'erita Universidad Aut\'onoma de Puebla.\\
\vskip 0.5cm
$^{(b)}${\it Facultad de Ciencias F\'{\i}sico-Matem\'aticas, BUAP.\\
Apdo. Postal 1364, C.P. 72000 Puebla, Pue., M\'exico}\\}
\vskip 0.5cm
$^{(c)}${\it Instituto de F\'{\i}sica, BUAP.\\
Apdo. Postal J-48, Col. San Manuel, C.P. 72570 Puebla, Pue., M\'exico}\\}
\end{center}
\vskip 0.5cm

\begin{abstract}
Predictive Higgs-fermion couplings can be obtained when a specific texture
for the fermion mass matrices is included in the general two-Higgs doublet
model. We derive the form of these couplings in the charged lepton sector
using a Hermitian mass matrix {\it Ansatz} with four-texture zeros.
The presence of unconstrained phases in the vertices $\phi_i l_i l_j$
modifies the pattern of flavor-violating Higgs interactions. Bounds on the
model parameters are obtained from present limits on rare lepton flavor
violating processes, which could be extended further by the search for
the decay $\tau \to \mu\mu\mu$ and $\mu-e$ conversion at future experiments.
The signal from Higgs boson decays $\phi_i \to \tau\mu$ could be searched at
the large hadron collider (LHC), while $e-\mu$ transitions could produce a
detectable signal at a future $e\mu$-collider, through the reaction
$e^+ \mu^- \to h^0 \to \tau^+ \tau^-$.

\bigskip

\end{abstract}

\end{titlepage}

\setcounter{footnote}{0}
\setcounter{page}{2}
\setcounter{section}{0}
\setcounter{subsection}{0}
\setcounter{subsubsection}{0}


\section{Introduction.}

After many years of success of the Standard Model (SM), the Higgs mechanism 
is still the least tested sector, and the problem of electroweak
symmetry breaking (EWSB) remmains almost as open as ever. However, the 
analysis of
raditive corrections within the SM \cite{hixradc}, points towards the existence
of a light Higgs boson, which could be detected in the early stages of LHC 
\cite{hixphen}. On the other hand, the SM is often considered 
as an effective theory, valid up to an energy scale of $O(TeV)$,
and eventually it will be replaced by a more fundamental theory,
which will explain, among other things, the physics behind EWSB 
and perhaps even the origin of flavor. Several examples of candidate 
theories, which range 
from supersymmetry \cite{susyrev} to deconstruction \cite{deconst},
include a Higgs sector with two scalar doublets, which has 
a rich structure and predicts interesting phenomenology \cite{mssmhix}.  
 The general two-higgs doublet model (THDM) has 
a potential problem with flavor changing neutral currents (FCNC)
mediated by the Higgs bosons, which arises when each quark type (u and d) 
is allowed to couple to both Higgs doublets, and FCNC could be induced at
large rates that may jeopardize the model. 
The possible solutions to this problem of the THDM 
involve an assumption about the Yukawa structure of the model. 
To discuss them it is convenient to refer to the Yukawa lagrangian, which is
written for the quarks fields as follows:
\begin{equation}
{\cal{L}}_Y = Y^{u}_1\bar{Q}_L \Phi_{1} u_{R} + 
                   Y^{u}_2 \bar{Q}_L\Phi_{2}u_{R} +
Y^{d}_1\bar{Q}_L \Phi_{1} d_{R} + Y^{d}_2 \bar{Q}_L\Phi_{2}d_{R} 
\end{equation}
where $\Phi_{1,2}=(\phi^+_{1,2}, \phi^0_{1,2})^T$ denote the Higgs doublets.
The specific choices for the Yukawa matrices $Y^q_{1,2}$ ($q=u,d$) define
the versions of the THDM known as I, II or III, which involve
the following mechanisms, that are aimed either to eliminate the otherwise
unbearable FCNC problem or at least to keep it under control, namely:

\begin{enumerate}
\item {\it{DISCRETE SYMMETRIES.}}
A discrete symmetry can be invoked to allow a given fermion
type (u or d-quarks for instance) to couple to a single Higgs
doublet, and in such case FCNC's are absent at tree-level. 
In particular, when a single Higgs field gives masses to both types 
of quarks (either $Y^u_1=Y^d_1=0$ or $Y^u_2=Y^d_2=0$), the resulting
model is referred as THDM-I. On the other hand, when each type of quark 
couples to a different Higgs doublet (either $Y^u_1=Y^d_2=0$ or 
$Y^u_2=Y^d_1=0$), the model is known as the THDM-II.
 This THDM-II pattern is highly motivated because it arises at tree-level 
in the minimal SUSY extension for the SM (MSSM) \cite{mssmhix}.

\item {\it{RADIATIVE SUPRESSION.}} 
When each fermion type couples to both Higgs doublets, 
FCNC could be kept under control if there exists
a hierarchy between $Y^{u,d}_1$ and $Y^{u,d}_2$.
Namely, a given set of Yukawa matrices is present at tree-level,
but the other ones  arise only as a radiative effect. 
This occurs for instance in the MSSM, where the 
type-II THDM structure is not protected by any symmetry, 
and is transformed into a type-III THDM (see bellow), through 
the loop effects of sfermions and gauginos. 
Namely, the Yukawa couplings that are already present at tree-level
in the MSSM ($Y^d_1, Y^u_2$) receive radiative corrections, while the terms
($Y^d_2, Y^u_1$) are induced at one-loop level. 

 In particular, when the ``seesaw'' mechanism~\cite{seesaw} is 
implemented in the MSSM to explain the observed neutrino 
masses ~\cite{atmospheric,solar}, lepton flavor violation (LFV) 
appears naturally in the right-handed neutrino sector, which is then 
communicated to the sleptons and from there to the charged leptons 
and Higgs sector. These corrections allow the neutral 
Higgs bosons to mediate LFV's, in particular  it was found 
that the (Higgs-mediated) tau decay $\tau \to 3\mu$ \cite{bakotau}
as well as the (real) Higgs decay $H\to \tau\mu$ \cite{myhlfvA}, 
can enter into possible detection domain. 
Similar effects are known to arise in the quark sector,
for instance $B\to\mu\mu$ can reach branching fractions 
at large $\tan\beta$, that can be probed at Run~II of the 
Tevatron~\cite{prlbako,bmuanalysis}. 

\item {\it{FLAVOR SYMMETRIES.}}
Suppression for FCNC can also be achived when a certain form of
the Yukawa matrices that reproduce the observed fermion masses and
mixing angles is implemented in the model, which is then named as 
THDM-III.  This could be done either by implementing the Frogart-Nielsen 
mechanism to generate the fermion mass hierarchies \cite{FN},
or by studying a certain {\it Ansatz} for the fermion mass matrices 
\cite{fritzsch}. The first proposal for the Higgs couplings
along these lines was posed
in \cite{chengsher,others}, it was based on the six-texture 
form of the mass matrices, namely:
\begin{displaymath}
M_l= 
\left( \begin{array}{ccc}
0 & C_{l} & 0 \\
C_{l}^{*} & 0 & B_{l} \\
0 & B_{l}^{*} & A_{l}
\end{array}\right).
\end{displaymath}
Then, by assuming that each Yukawa matrix $Y^q_{1,2}$ has the same 
hierarchy, one finds: $A_{l}\simeq m_3$, $B_{l}\simeq \sqrt{m_2m_3}$ and 
$C_{l}\simeq \sqrt{m_1m_2}$. Then, the Higgs-fermion couplings 
obey the following pattern:
$Hf_{i}f_{j} \sim \sqrt{m_{i}m_{j}} / m_{W}$, 
which is known as the Cheng-Sher {\it Ansatz}. This brings under control
the FCNC problem, and it has been extensively studied in the literature to
search for flavor-violating signals in the Higgs sector \cite{muchos}.

\end{enumerate}

In this paper we are interested in studying the flavor symmetry option.
However, the six-texture {\it Ansatz} seems disfavored by current data
on the CKM mixing angles. More recently, mass matrices with
four-texture {\it Ansatz} have been considered, and are found in better 
agreement with the observed data \cite{fourtext}. It is interesting
then to investigate how the Cheng-Sher form of the Higgs-fermion 
couplings, gets modified when one replaces the six-texture matrices 
by the four-texture {\it Ansatz}. This paper is aimed precisely to study
this question; we want to derive the form of the Higgs-fermion couplings
and to discuss how and when the resulting predictions could be 
tested, both in rare tau decays and in the phenomenology of 
the Higgs bosons \cite{myhlfvA}. Unlike previous studies, we
keep in our analysis the effect of the complex phases, which modify the
FCNC Higgs couplings.

The organization of the paper goes as follows: In section 2, we discuss the
lagrangian for the THDM with the four-texture form for the mass matrices,
and present the results for the Higgs-fermion vertices in the charged lepton
sector. Then, in section 3 we study the constraints impossed on the
parameters of the model from low energy LFV processes. In section 4 we
discuss predictions of the model for tau and Higgs decays, including the
capabilities of future hadron and $e\mu$-colliders to probe this phenomena.
Finally, section 5 contains our conclusions.

\section{The THDM-III with four-texture mass matrices}

The Yukawa lagrangian of the THDM-III
for the lepton sector is given by:

\begin{equation}
{\cal{L}}_Y^l = Y^{l}_{1ij}\bar{L_{i}}\Phi_{1}l_{Rj} + 
Y^{l}_{2ij}\bar{L_{i}}\Phi_{2}l_{Rj}.
\end{equation}

After SSB the charged lepton mass matrix is given by,
\begin{equation}
M_l= \frac{1}{\sqrt{2}}(v_{1}Y_{1}^{l}+v_{2}Y_{2}^{l}),
\end{equation}

We shall asuume that both Yukawa matrices $Y^l_1$ and $Y^l_2$ have the
four-texture form and Hermitic; following the conventions of \cite{fourtext}, the lepton mass matrix
is then written as:

\begin{displaymath}
M_l= 
\left( \begin{array}{ccc}
0 & C_{l} & 0 \\
C_{l}^{*} & \tilde{B}_{l} & B_{l} \\
0 & B_{l}^{*} & A_{l}
\end{array}\right).
\end{displaymath}
when $\tilde{B}_{l}\to 0$ one recovers the six-texture form.
We also consider the hierarchy: \\
$\mid A_{l}\mid \, \gg \, \mid \tilde{B}_{l}\mid,\mid B_{l}\mid ,\mid C_{l}\mid$,
which is supported by the observed fermion masses in the SM.

Because of the hermicity condition, both $\tilde{B}_{l}$ and
$A_{l}$ are real parameters, while the phases of $C_l$ and $B_l$,
$\Phi_{B,C}$, can be removed from the mass matrix $M_l$
by defining: $M_l=P^\dagger \tilde{M} P$, where 
$P=diag[1, e^{i\Phi_C},  e^{i(\Phi_B+\Phi_C)}]$, 
and the mass matrix $\tilde{M}_l$ includes only the real parts
of $M_l$.
The  diagonalization of $\tilde{M}$ is then obtained by an orthogonal matrix 
$O$, such that the diagonal mass matrix is: 
$\bar{M}_{l} = O^{T}\tilde{M}_{l}O$.

\bigskip

The lagrangian (2) can be expanded in terms of
the mass-eigenstates for the neutral ($h^0,H^0,A^0$) and 
charged Higgs bosons ($H^\pm$). The
interactions of the neutral Higgs bosons are given by,

\begin{eqnarray}
{\cal{L}}_Y^{l} & = & \frac{g}{2}\left(\frac{m_{i}}{m_W}\right)
\bar{l}_{i}\left[\frac{ \, \cos\alpha}{\cos\beta}\delta_{ij}+ 
\frac{\sqrt{2} \, \sin(\alpha - \beta)}{g \, \cos\beta}
\left(\frac{m_W}{m_{i}}\right)\tilde{Y}_{2ij}^{l}\right]l_{j}H^{0} 
\nonumber \\
                 &  &+ \frac{g}{2}\left(\frac{m_{i}}{m_W}\right)\bar{l}_{i}
\left[-\frac{\sin\alpha}{\cos\beta} \delta_{ij}+  
\frac{\sqrt{2} \, \cos(\alpha - \beta)}{g \, \cos\beta}
\left(\frac{m_W}{m_{i}}\right)\tilde{Y}_{2ij}^{l}\right]l_{j} h^{0}
\nonumber \\
                 & &+ \frac{ig}{2}\left(\frac{m_{i}}{m_W}\right)\bar{l}_{i}
\left[-\tan\beta \delta_{ij}+  \frac{\sqrt{2} }{g \, \cos\beta}
\left(\frac{m_W}{m_{i}}\right)\tilde{Y}_{2ij}^{l}\right]
\gamma^{5}}l_{j} A^{0.
\end{eqnarray}
The first term, proportional to $\delta{ij}$ corresponds to the
modification of the THDM-II over the SM result, while the term
proportional to $\tilde{Y}_2^l$ denotes the new contribution from
THDM-III. Thus, the fermion-Higgs couplings respect CP-invariance,
despite the fact that the Yukawa matrices include complex phases;
this follows because of the Hermiticity conditions impossed on both
$Y_1^l$ and $Y_2^l$.

The corrections to the lepton flavor conserving (LFC) and flavor-violating
(LFV) couplings, depend on the rotated matrix:  $\tilde{Y}_{2}^{l} =
O^{T}PY_{2}^{l}P^\dagger O$. We shall evaluate $\tilde{Y}_{2}^{l}$, by
assuming that $Y_2^l$ has a four-texture form, namely:

\begin{equation}
Y_{2}^{l}  =
\left( \begin{array}{ccc}
0 & C_{2} & 0 \\
C_{2}^{*} & \tilde{B}_{2} & B_{2} \\
0 & B_{2}^{*} & A_{2}
\end{array}\right), \qquad
\mid A_{2}\mid \, \gg \, \mid \tilde{B}_{2}\mid,\mid B_{2}\mid ,\mid C_{2}\mid.
\end{equation}

The matrix that diagonalizes the real matrix
$\tilde{M}_{l}$ with the four-texture form, is given by:

\begin{displaymath}
O =
\left( \begin{array}{ccc}
\sqrt{\frac{\lambda_{2}\lambda_{3}(A-\lambda_{1})}{A(\lambda_{2}-\lambda_{1})
(\lambda_{3}-\lambda_{1})}}& \eta \sqrt{\frac{\lambda_{1}\lambda_{3}
(\lambda_{2}-A)}{A(\lambda_{2}-\lambda_{1})(\lambda_{3}-\lambda_{2})}}
& \sqrt{\frac{\lambda_{1}\lambda_{2}(A-\lambda_{3})}{A(\lambda_{3}-
\lambda_{1})(\lambda_{3}-\lambda_{2})}} \\
-\eta \sqrt{\frac{\lambda_{1}(\lambda_{1}-A)}{(\lambda_{2}-\lambda_{1})
(\lambda_{3}-\lambda_{1})}} & \sqrt{\frac{\lambda_{2}(A-\lambda_{2})}
{(\lambda_{2}-\lambda_{1})(\lambda_{3}-\lambda_{2})}} & \sqrt{
\frac{\lambda_{3}(\lambda_{3}-A)}{(\lambda_{3}-\lambda_{1})(\lambda_{3}-
\lambda_{2})}} \\
\eta \sqrt{\frac{\lambda_{1}(A-\lambda_{2})(A-\lambda_{3})}{A(\lambda_{2}
-\lambda_{1})(\lambda_{3}-\lambda_{1})}} & -\sqrt{\frac{\lambda_{2}(A
-\lambda_{1})(\lambda_{3}-A)}{A(\lambda_{2}-\lambda_{1})(\lambda_{3}
-\lambda_{2})}} & \sqrt{\frac{\lambda_{3}(A-\lambda_{1})(A-\lambda_{2})}
{A(\lambda_{3}-\lambda_{1})(\lambda_{3}-\lambda_{2})}}
\end{array}\right),
\end{displaymath}
where $m_{e}= m_{1} = \mid \lambda _{1}\mid, m_{\mu}= m_{2} = \mid
\lambda _{2}\mid,
m_{\tau}= m_{3} = \mid \lambda _{3}\mid, \eta = \lambda_{2}/ m_{2} $
  
Then the rotated form $\tilde {Y}_{2}^{l} $ has the general form,

\begin{eqnarray}
\tilde {Y}_{2}^{l}  & = & O^{T}PY_{2}^{l}P^{\dagger}O \nonumber \\
& = &\left( \begin{array}{ccc}
\tilde {Y}_{211}^{l}   & \tilde {Y}_{212}^{l}   & \tilde {Y}_{213}^{l}   \\
\tilde {Y}_{221}^{l}   & \tilde {Y}_{222}^{l}   & \tilde {Y}_{223}^{l}  \\
\tilde {Y}_{231}^{l}   & \tilde {Y}_{232}^{l}   & \tilde {Y}_{233}^{l}
\end{array}\right).
\end{eqnarray}

However, the full expressions for the resulting elements have a complicated 
form, as it can be appreciated, for instance, by looking at the element  
$(\tilde{Y}_{2}^{l})_{22}$, which is displayed here:

\begin{eqnarray}
(\tilde{Y}_{2})_{22}^{l} &=& \eta [C^*_2 e^{i\Phi_C} +C_2 e^{-i\Phi_C}]
\frac{(A-\lambda_{2})}{m_3-\lambda_2 } \sqrt{\frac{m_1 m_3 }{A m_2}} +
 \tilde{B}_2 \frac{A-\lambda_2}{ m_3-\lambda_2 } \\
& & + A_2 \frac{A-\lambda_2}{ m_3-\lambda_2 }
- [B^*_2 e^{i\Phi_B} + B_2 e^{-i\Phi_B}]
\sqrt{\frac{(A-\lambda_{2})(m_3-A) } {m_3- \lambda_2}} 
\end{eqnarray}
where we have taken the limits: $ |A|,m_{\tau},m_{\mu} \gg m_{e}$.
The free-parameters are: $\tilde{B_{2}}, B_{2}, A_{2}, A$. 

To derive a better suited approximation, we shall consider the elements of
the Yukawa matrix $Y_2^l$ as having the same hierarchy as the full mass
matrix, namely:
 
\begin{eqnarray}
C_{2} & = &  c_{2}\sqrt{\frac{m_{1}m_{2}m_{3}}{A}}  \\
B_{2} & = &  b_{2}\sqrt{(A - \lambda_{2})(m_{3}-A)}  \\
\tilde{B}_{2} & = & \tilde{b}_{2}(m_{3}-A + \lambda_{2})  \\
A_{2} & = & a_{2}A.
\end{eqnarray}

Then, in order to keep the same hierarchy for the elements of the mass 
matrix, we find that $A$ must fall within the interval $ (m_{3}- m_{2})
\leq A \leq m_{3}$. Thus, we propose the following relation for $A$:

\begin{equation}
A  = m_{3}(1 -\beta z),
\end{equation} 
where $z = m_{2}/m_{3} \ll 1$  and $0 \leq \beta \leq 1$.

Then, we introduce the matrix $\tilde{\chi}$ as follows:

\begin{eqnarray}
\left( \tilde {Y}_{2}^{l} \right)_{ij}
&=& \frac{\sqrt{m_i m_j}}{v} \, \tilde{\chi}_{ij} \nonumber\\
&=&\frac{\sqrt{m_i m_j}}{v}\, {\chi}_{ij} \, e^{\vartheta_{ij}}
\end{eqnarray}
which differs from the usual Cheng-Sher $Ansatz$ not only because of the
appearence of the complex phases, but also in the form of the real parts
${\chi}_{ij} = |\tilde{\chi}_{ij}|$.

Expanding in powers of $z$, one finds that the elements of the matrix
$\tilde{\chi}$ have the following general expressions:

\begin{eqnarray}
\tilde{\chi}_{11} & = &  
[\tilde{b}_{2}-(c^*_{2}e^{i\Phi_{C}} +c_{2}e^{-i\Phi_{C}} )]\eta 
    +[a_{2}+\tilde{b}_{2}-(b^*_{2}e^{i\Phi_{B}} + b_{2}e^{-i\Phi_{B}} )]
         \beta \nonumber \\
\tilde{\chi}_{12} & = & (c_{2}e^{-i\Phi_{C}}-\tilde{b}_{2}) -\eta[a_{2}+
\tilde{b}_{2}-(b^*_{2}e^{i\Phi_{B}} + b_{2}e^{-i\Phi_{B}} )] \beta ]
\nonumber \\
\tilde{\chi}_{13} & = & (a_{2}-b_{2}e^{-i\Phi_{B}}) \eta \sqrt{\beta}
                           \nonumber  \\
\tilde{\chi}_{22}  & = & \tilde{b}_{2}\eta
+[a_{2}+\tilde{b}_{2}-(b^*_{2}e^{i\Phi_{B}} +b_{2}e^{-i\Phi_{B}} )]
         \beta \nonumber \\
\tilde{\chi}_{23} & = & (b_{2}e^{-i\Phi_{B}}-a_{2})
                              \sqrt{\beta} \nonumber  \\
\tilde{\chi}_{33} & = & a_{2}
\end{eqnarray}

\noindent It is also relevant to point out the following:

\begin{itemize}
\item When the phases $\Phi_B$ and $\Phi_C$ vanish, $\beta = 1$ and
one takes the 6-texture limit ($\tilde{B_2} \to 0, i.e. \, \tilde{b} \to 0
 \Rightarrow \eta =-1$), Eq. (14) reduces to

\begin{eqnarray}
\left( \tilde {Y}_{2}^{l} \right)_{11} & = &  
(2 c_2 + a_2 -2 b_2) \, m_1/v \nonumber \\
\left( \tilde {Y}_{2}^{l} \right)_{12} & = &
(c_2 + a_2 -2 b_2) \, \sqrt{m_1 m_2}/v \nonumber \\
\left( \tilde {Y}_{2}^{l} \right)_{13} & = &
(b_2 - a_2) \, \sqrt{m_1 m_3}/v \nonumber \\
\left( \tilde {Y}_{2}^{l} \right)_{22} & = & 
(a_2 -2 b_2) \, m_2/v \nonumber \\
\left( \tilde {Y}_{2}^{l} \right)_{23} & = &
(b_2 - a_2) \, \sqrt{m_2 m_3}/v \nonumber \\
\left( \tilde {Y}_{2}^{l} \right)_{33} & = & a_{2} \, m_3/v
\end{eqnarray}

which correspond to the {\it Ansatz} of Cheng-Sher (See Eq. (32) in
Ref. \cite{chengsher}).

\item On the other hand, when the phases $\Phi_B$ and $\Phi_C$ vanish,
$\beta = m_2/m_3$ and $\eta = 1$, Eq. (14) reduces to

\begin{eqnarray}
\left( \tilde {Y}_{2}^{l} \right)_{11} & = &  
(\tilde{b}_2 - 2 c_2) \, m_1/v \nonumber \\
\left( \tilde {Y}_{2}^{l} \right)_{12} & = &
(c_2 - \tilde{b}_2) \, \sqrt{m_1 m_2}/v \nonumber \\
\left( \tilde {Y}_{2}^{l} \right)_{13} & = &
(a_2 - b_2) \, \sqrt{m_1 m_2}/v \nonumber \\
\left( \tilde {Y}_{2}^{l} \right)_{22} & = & 
\tilde{b}_2 \, m_2/v \nonumber \\
\left( \tilde {Y}_{2}^{l} \right)_{23} & = &
(b_2 - a_2) \, m_2/v \nonumber \\
\left( \tilde {Y}_{2}^{l} \right)_{33} & = & a_{2} \, m_3/v
\end{eqnarray}

in this case one reproduces the results given in Ref.\cite{zhou}
(See Eq. (24)).

\end{itemize}

\bigskip

While the diagonal elements $\tilde{\chi}_{ii}$ are real, we notice
(Eqs. 15) the appearance of the phases in the off-diagonal elements,
which are essentially unconstrained by present low-energy phenomena.
As we will see next, these phases modify the pattern of flavor violation
in the Higgs sector. For instance, while the Cheng-Sher $Ansatz$ predicts
that the LFV couplings $(\tilde{Y}_2^l)_{13}$ and $(\tilde{Y}_2^l)_{23}$
vanish when $a_2 = b_2$, in our case this is no longer valid for
$\cos\Phi_B \neq 1$. Furthermore the LFV couplings satisfy several
relations, such as: $|\tilde{\chi}_{23}| = |\tilde{\chi}_{13}|$,
which simplifies the parameter freedom.

Finally, in order to perform our phenomenological study we find convenient
to rewrite the lagrangian given in Eq. (4) in terms of the
$\tilde{\chi}_{ij}$'s as follows:

\begin{eqnarray}
{\cal{L}}_Y^{l} & = & \frac{g}{2} \, \bar{l}_{i}
\left[\left( \, \frac{m_{i}}{m_W}\right)\frac{\cos\alpha}{\cos\beta} \,
\delta_{ij} + \frac{\sin(\alpha - \beta)}{\sqrt{2} \, \cos\beta}
\left(\frac{\sqrt{m_i m_j}}{m_W}\right)\tilde{\chi}_{ij}\right]l_{j}H^{0} 
\nonumber \\
                &   & + \frac{g}{2} \, \bar{l}_{i}
\left[-\left(\frac{m_{i}}{m_W}\right)\frac{\sin\alpha}{\cos\beta} \,
\delta_{ij} + \frac{\cos(\alpha - \beta)}{\sqrt{2} \, \cos\beta}
\left(\frac{\sqrt{m_i m_j}}{m_W}\right)\tilde{\chi}_{ij}\right]l_{j} h^{0}
\nonumber \\
                &   & + \frac{ig}{2} \, \bar{l}_{i}
\left[-\left(\frac{m_{i}}{m_W}\right)\tan\beta \, \delta_{ij} +
\frac{1}{\sqrt{2} \, \cos\beta}
\left(\frac{\sqrt{m_i m_j}}{m_W}\right)\tilde{\chi}_{ij}\right]
\gamma^{5}}l_{j} A^{0.
\end{eqnarray}
where, unlike the Cheng-Sher {\it Ansatz}, $\tilde{\chi}_{ij}$ $(i \neq j)$
are complex.

\section{Bounds on the LFV Higgs parameters}

Constrains on the LFV-Higgs interaction will be obtained by studying
LFV transitions, which include the 3-body modes 
($l_i \to l_j l_k \bar{l}_k$), radiative decays ($l_i \to l_j +\gamma$),
$\mu-e$ conversion in nuclei, as well as the (LFC) muon anomalous
magnetic moment.

\bigskip

\noindent 3.1 {\it LFV three-body decays}. To evaluate the LFV leptonic
couplings, we calculate the decays
$l_{i} \to l_j l_k \bar{l}_k$, including the contribution from
the three Higgs bosons ($h^0$, $H^0$ and $A^0$).We obtain the following
expression for the branching ratio:

\begin{eqnarray}
Br (l_{i} \to l_j l_k \bar{l}_k) &=& 
\frac{5 \, \delta_{jk} + 2}{3} \, \frac{\tau_i}{2^{11} \, \pi^3} \,
\frac{m_j \, m_k^2 \, m_i^6}{v^4} \,
\left \{ \frac{\cos^2(\alpha-\beta) \, \sin^2\alpha}{m^4_{h^0}}
+ \frac{\sin^2(\alpha-\beta) \, \cos^2\alpha}{m^4_{H^0}} \right. \nonumber\\
& & -2 \, \frac{\cos(\alpha-\beta) \, \sin(\alpha-\beta) \, \cos\alpha \,
\sin\alpha}{m^2_{h^0} \, m^2_{H^0}} + \left. \frac{\sin^2\beta}{m^4_{A^0}}
\right \} \frac{\chi_{ij}^2}{2 \, \cos^4\beta}
\end{eqnarray}
where $\tau_i$ denotes the life time of the lepton $l_i$ and we have
assumed $\chi_{kk} \ll 1$; this result agrees with Ref. \cite{zhou}.

In particular, for the decay
$\tau^- \to \mu^- \mu^+ \mu^-$ we obtain the following expression for the
branching ratio:

\begin{eqnarray}
Br (\tau^- \to \mu^- \mu^+ \mu^-) &=& 
\frac{5}{3} \, \frac{\tau_{\tau}}{2^{12} \, \pi^3} \,
\frac{m_2^3 \, m_3^6}{v^4} \,
\left \{ \frac{\cos^2(\alpha-\beta) \, \sin^2\alpha}{m^4_{h^0}}
+ \frac{\sin^2(\alpha-\beta) \, \cos^2\alpha}{m^4_{H^0}} \right. \nonumber\\
& & -2 \, \frac{\cos(\alpha-\beta) \, \sin(\alpha-\beta) \, \cos\alpha \,
\sin\alpha}{m^2_{h^0} \, m^2_{H^0}} + \left. \frac{\sin^2\beta}{m^4_{A^0}}
\right \} \frac{\chi_{23}^2}{\cos^4\beta}
\end{eqnarray}
here $\tau_{\tau}$ corresponds to the life time of the $\tau$ lepton
(we have also assumed $\chi_{22} \ll 1$).

Using the experimental result $Br(\tau^- \to \mu^- \mu^+ \mu^-) <
1.9 \times 10^{-6}$, we get an upper bound on $\chi_{23}$
($(\chi_{23})_{u.\,b.}^{\tau \to 3\mu}$) as a function of
$\alpha$ and $\tan\beta$. In Fig. 1 we show the value of this bound as a
function of $\tan\beta$ for $\alpha = \beta - \pi /4$,
$\alpha = \beta - \pi /3$ and $\alpha = \beta - \pi /2$,
taking $m_{h^0} = 115$ $GeV$ and $m_{H^0} = m_{A^0} = 300$ $GeV$.

Taking $\chi_{23} \approx 1$, $\tan\beta \approx 30$ and
$\pi/4 < \beta - \alpha < \pi/2$, in Eq. (20)
one finds tipically that $Br(\tau\to3\mu) \sim 10^{-8}$,
which puts it into the regime that is experimentally accessible
at $\tau$-factories over the next few years. At LHC and SuperKEKB,
limits in the range of $10^{-9}$ should be achievable \cite{exp},
allowing a deeper probe into the parameter space.

\bigskip

\noindent 3.2 {\it Radiative decays}. The branching ratio of
$\mu^+ \to e^+ \gamma$ at one loop level is given by \cite{chang1}

\begin{eqnarray}
Br(\mu^+ \to e^+ \gamma) &=& \frac{\alpha_{em} \tau_{\mu} m_1 m^4_2 m^4_3}
{2^{12} \pi^4 v^4 \cos^4 \beta} \, \chi_{23}^2 \, \chi_{13}^2
\left \{ \frac{\cos^4(\alpha-\beta)}{m^4_{h^0}}
 \left| \ln \frac{m^2_{3}}{m^2_{h^0}}+\frac{3}
{2} \right|^2 \right. \nonumber\\
&&+2 \frac{\cos^2(\alpha-\beta)\sin^2(\alpha-\beta)}{m^2_{h^0}m^2_{H^0}}
 \left| \ln \frac{m^2_{3}}{m^2_{h^0}}+\frac{3}
{2} \right| 
 \left| \ln \frac{m^2_{3}}{m^2_{H^0}}+\frac{3}
{2} \right| \nonumber\\
&&+\left. \frac{\sin^4(\alpha-\beta)}{m^4_{H^0}}
 \left| \ln \frac{m^2_{3}}{m^2_{H^0}}+\frac{3}
{2} \right|^2
+ \frac{1}{m^4_{A^0}}
 \left| \ln \frac{m^2_{3}}{m^2_{A^0}}+\frac{3}
{2} \right|^2 \right \} 
\end{eqnarray}
From Eqs. (15) we have $\chi_{23} = \chi_{13} =
|(a_2 - b_2 e^{-i\Phi_B})|\sqrt{\beta}$. We will make use of the current
experimental upper bound $Br(\mu^+ \to e^+ \gamma) < 1.2 \times 10^{-11}$
\cite{partdata} to constraint $\chi_{23}
(\chi_{13})$ as a function of
$\alpha$ and $\tan\beta$. Assuming $m_{h^0} = 115$ $GeV$ and $m_{H^0} =
m_{A^0} = 300$ $GeV$, we depict in Fig. 2 the value of the upper bound
on $\chi_{23}$ ($(\chi_{23})_{u.\,b.}^{\mu \to e \gamma}$) as a
function of $\tan\beta$, again for $\alpha = \beta - \pi /4$,
$\alpha = \beta - \pi /3$ and $\alpha = \beta - \pi /2$.
A new experiment at PSI will measure the process $\mu^+ \to e^+ \gamma$
with a sensitivity of 1 event for $Br(\mu^+ \to e^+ \gamma) = 10^{-14}$
\cite{psiexp}, which would improve the upper bound on $\chi_{23}$ by a factor
$\sim 10^{-3/4} \approx 0.18$ .

\bigskip

\noindent 3.3 {\it $\mu-e$ conversion}. The formulas of the conversion
branching ratios for the lepton flavor violating muon electron process in
nuclei at large $\tan \beta$, in the aluminum and lead targets, are
approximately given by

\begin{equation}
Br(\mu^- Al \to e^- Al) \simeq 1.8 \times 10^{-4} \,
\frac{m_1 \, m^6_2 \, m^2_p \, \tan^6\beta \, \cos^2\beta}
{2 \, v^4 \, m^4_{H^0} \, \omega_{capt}}
\, \chi_{12}^2
\end{equation}
and
\begin{equation}
Br(\mu^- Pb \to e^- Pb) \simeq 2.5 \times 10^{-3} \,
\frac{m_1 \, m^6_2 \, m^2_p \, \tan^6\beta \, \cos^2\beta}
{2 \, v^4 \, m^4_{H^0} \, \omega_{capt}}
\, \chi_{12}^2,
\end{equation}
respectively, where $\omega_{capt}$ is the rate for muon capture in the nuclei
\cite{kitano}. The values are $\omega_{capt}=0.7054 \times 10^6 \, s^{-1}$
and $\omega_{capt}=13.45 \times 10^6 \, s^{-1}$ in the aluminum and the lead
nuclei, respectively \cite{suzuki}. There are several planned experiments which
are aiming at improving the bounds of the branching fractions for relevant
processes by three or four orders of magnitude \cite{barkov,bachman,kuno}. In particular, the
MECO experiment will search for the coherent conversion of muons to electrons
in the field of a nucleus with a sensitivity of 1 event for $5 \times 10^{16}$
muon captures, {\it i. e.} $Br(\mu^- {\cal N} \to e^- {\cal N}) < 2 \times
10^{-17}$ \cite{kosmas,molzon}. Taking $m_{H^0} = 300$ $GeV$, we plot in
Fig. 3 the value of the upper bound on $\chi_{12}$ ($(\chi_{12})_{u.\,b.}^
{\mu {\cal N} \to e {\cal N}}$) as a function of $\tan\beta$ for $Al$ and
$Pb$, for the current experimental measurement $Br(\mu^- {\cal N} \to e^-
{\cal N}) < 6.1 \times 10^{-13}$ \cite{molzon}. In Fig. 4, we show the same
as in Fig. 3 but for $Br(\mu^- {\cal N} \to e^- {\cal N}) < 2 \times
10^{-17}$.

\bigskip

\noindent 3.4 {\it Muon anomalous magnetic moment}. Taking the average
value of the measurements of the muon $(g-2)$ from \cite{bennett} and
the recent analysis by different groups \cite{amutheo1,amutheo2} one can
conlude that

\begin{equation}
\Delta a_{\mu} \equiv a_{\mu}^{exp} - a_{\mu}^{SM} \approx 300
\pm 100 \times 10^{-11}.
\end{equation}

The contribution to the
muon $g-2$ of the one loop level flavor changing diagram is given as follows.

\begin{equation}
\Delta a_{\mu} = {\pm} \frac{1}{16 \pi^2} \,
\frac{m^2_2 \, m^2_3}{v^2 \, m^2_{\phi^0}} \, 
\frac{\cos^2(\alpha-\beta)}{\cos^2\beta} \,
\left( \ln \frac{m^2_{\phi^0}}{m^2_3} -\frac{3}{2} \right) \,
\chi_{23}^2\\
\end{equation}
where the sign +(-) is for scalar $\phi^0=h^0$ (pseudoscalar, $\phi^0=A^0$)
exchanges \cite{zhou-wu,amu1,amu2,amu3}. From Eq. (24) is clear that we need
to increase the theoretical value of $a_{\mu}$. Hence, we will consider the
contribution of $h^0$ to the muon $(g-2)$ assuming $\chi_{23}=1$, namely
$\Delta a_{\mu}^{h^0}(\chi_{23}=1)$. We take $m_{h^0} = 115$ $GeV$ and
present in Fig. 5 the result
for such contribution as a function of $\tan\beta$ for $\alpha = \beta -
\pi /4$ and $\alpha = \beta - \pi /3$. We observe that 
$\Delta a_{\mu}^{h^0}(\chi_{23}=1) < 240 \times 10^{-11}$.
On the other hand, the contribution to the anomalous magnetic moment
from two-loop double scalar-exchanging diagrams is comparable with the one from
the corresponding flavor-changing one loop diagrams \cite{zhou-wu}. It was
already shown in Ref. \cite{zhou-wu} that the two-loop double scalar
(pseudoscalar) exchanging diagrams give negative (positive) contributions,
which have opposite signs as the one from one loop scalar (pseudoscalar)
exchanging diagram. Hence, we can conclude that it would be very hard
to constrain $\chi_{23}$ from the muon $g-2$ measurements, or to explain
such deviation from the pure Higgs sector in case the signal is confirmed.

\bigskip

Thus, we conclude from this section that the bounds on the LFV parameters
are given as follows.
\begin{itemize}
\item $\chi_{12} < 5 \times 10^{-1}$, from $\mu^--e^- $ conversion
experiments.

\item $\chi_{13} = \chi_{23} < 6 \times 10^{-1}$,
from the radiative decay $\mu^+ \to e^+ \gamma$ measurements.
\end{itemize}

\noindent However, one can still say that at the present time the
couplings $\chi_{ij}$'s are not highly constrained, thus they could induce
interesting direct LFV Higgs signals at future colliders.

\section{Probing the LFV Higgs couplings at future colliders}

In order to probe the LFV Higgs vertices we shall consider both
the search for the LFV Higgs decays at future hadron colliders
(LHC mainly), as well as the production of Higgs bosons in the
collisions of electrons and muons, which was proposed some time
ago \cite{emucoll}, namely we shall evaluate the reaction $e\mu
\to h^0 \to \tau \tau$.

\bigskip

\noindent 4.1 {\it Search for LFV Higgs decays at Hadron colliders.}
We shall concentrate here on the LFV Higgs decays $\phi_i \to \tau \mu$, 
which has a very small  branching ratio within the context of the SM with
light neutrinos ($\leq 10^{-7}-10^{-8}$ ), so that this channel becomes 
an excellent window for probing new physics\,\cite{myhlfvA,myhlfvB,otherlfvh}.
The decay width for the procces $\phi_i \to \tau \mu$ 
(adding both final states $\tau^+ \mu^-$ and $\tau^- \mu^+$ )
can be written in terms of the decay width $\Gamma (H_i \to \tau \tau)$,
as follows:
\beq
\Gamma (\phi_i\to \tau \mu) \, =\, (R^{\,\phi}_{\tau\mu})^2
                                   \, \Gamma (H_i \to \tau \tau)
\eeq
where
\beq
R^{\,\phi}_{\tau\mu}=\frac{g_{\phi \tau \mu}}{g_{\phi \tau \tau}}
\cong \frac{\sin(\alpha-\beta)}{\cos\alpha} \sqrt{\frac{m_{\mu}}{m_{\tau}}}
\, \tilde{\chi}_{23}
\eeq
Therefore, the Higgs branching ration can be approximated as:
$Br(\phi_i \to \tau \mu)= (R^{\,\phi}_{\tau\mu})^2 \times 
Br(\phi_i \to \tau \tau)$. We calculated the branching fraction for
$h \to \tau\mu$,  and find that it reaches
values of order $10^{-2}$ in the THDM-III; for comparison, we notice
that in the MSSM case, even for large values of $\tan\beta$, 
one only gets  $Br(h \to \tau\mu)\simeq 10^{-4}$.

These values of the branching ratio enter into the domain of
detectability at hadron colliders (LHC), provided that the cross-section
for Higgs production were of order of the SM one.  Large values of
$\tan\beta$ are also associated with large b-quark Yukawa coupling,
which in turn can produce and enhancement on the Higgs production
cross-sections at hadron colliders, even for the heavier states $H^0$ and
$A^0$ either by gluon fusion or in the associated production of the Higgs
with b-quark pairs; some values are shown in table 1; these were obtained
using HIGLU \cite{hspira}. Thus, even the heavy Higgs bosons of the model
could be detected through this LFV mode.

\bigskip

\begin{center}
\begin{tabular}{|c|c|c|c|}
\hline
$m_{H,A}$ [GeV] & $\sigma^H_{gg}$ [pb] & $\sigma^A_{gg}$ [pb] &
 $\sigma^H_{bb}$ [pb] ($\simeq \sigma^A_{bb}$) \\
\hline
 150   & 126.4 (492.6) & 129.1 (525.) & 200 (800) \\
\hline
 200   & 29.5  (114.3) & 29.1 (120.)  & 100 (400) \\
\hline
 300   & 3.6   (13.5)  & 3.15 (13.6)  & 20 (80)  \\
\hline
 350   & 1.6   (5.9)   & 1.2  (5.6)   & 12 (48)  \\
\hline
 400   & 0.75  (2.75)  & 0.73 (2.8)   & 8 (32)  \\
\hline
\end{tabular}
\end{center}
\noindent
{Table\,1. Cross-section for Higgs production at LHC, through gluon
fusion ($\sigma^{H,A}_{gg}$) and in association with $b\bar{b}$ quarks,
($\sigma^{H,A}_{bb}$), for $\tan\beta=30$ (60).}

\bigskip

For instance, for $m_{H,A}=150$ GeV and $\tan\beta=30 (60)$
the cross-section through gluon fusion at LHC is about 
126.4 (492.6) pb \cite{hspira}, then with 
$Br(H\to \tau \mu) \simeq 10^{-2} (10^{-3})$ and
an integrated luminosity of $10^{5} \, pb^{-1}$,
LHC can produce about $10^5 (10^4)$ LFV Higgs events.
In Ref. \cite{htaumubkd} it was proposed a series of cuts to reconstruct
the hadronic and electronic tau decays from $h\to \tau\mu$ and separate the
signal from the backgrounds, which are dominated by Drell-Yan tau pair
and WW pair production.
According to these studies \cite{htaumubkd}, even SM-like cross 
sections and $m_\phi \simeq 150 \, GeV$, one coud detect at LHC
the LFV Higgs decays with a branching ratio of order $8 \times 10^{-4}$,
which means that our signal is clearly detectable.

\bigskip

\noindent 4.2 {\it Tests of LFV Higgs couplings at $e\mu$-colliders.}
Another option to search for LFV Higgs couplings, but
now involving the electron-muon-Higgs couplings, would be
to search for the reaction:
$e^-(p_a) + \mu^+(p_b) \to h^0 \to \tau^-(p_c) + \tau^+(p_d)$.
Assuming $\chi_{33} \ll 1$, the result for the cross section
is given by:
\begin{eqnarray}
\sigma(e^- \mu^+ \to \tau^- \tau^+)&=&\frac{s \, m_1 \, m_2 \, m_3^2 \,}
{32 \, \pi \, v^4\, \cos^4\beta}
 \, \chi_{12}^2 \nonumber\\
 && \{ |D_{h^0}(s)|^2 \cos^2(\alpha-\beta) \, \sin^2\alpha \nonumber\\
 && -2 Re \{D_{h^0}(s) \, D^{\star}_{H^0}(s) \} \, \cos(\alpha-\beta)
      \,\sin(\alpha-\beta) \, \sin\alpha \, \cos\alpha\nonumber\\
 &&   \, +|D_{H^0}(s)|^2 \sin^2(\alpha-\beta) \, \cos^2\alpha +
      |D_{A^0}(s)|^2 \, \sin^2\beta \},
\end{eqnarray}
where $D_{\phi^0}(s)$ denotes the Breit-Wigner form of the $\phi^0$ propagator
\begin{equation}
D_{\phi^0}(s) = (s-m^2_{\phi^0}+i m_{\phi^0}\,\Gamma^{\phi^0}_{tot})^{-1}
\end{equation}
and $s=(p_a+p_b)^2=(p_c+p_d)^2$.

The non-observation of at least an event in a year would imply that
\begin{equation}
\sigma(e^- \mu^+ \to \tau^- \tau^+) \times \mbox{luminosity} \times 1
\, \mbox{year} < 1,
\end{equation}
which would allow us to put an upper bound on $\chi_{12}$,
namely $(\chi_{12})_{u.\,b.}^{e\mu\to\tau\tau}(s)$
as a function of $\alpha$ and $\tan\beta$.
In order to obtain numerical results, we take $\Gamma^{h^0}_{tot}=0.004
\, GeV$ for $m_{h^0}=115 \, GeV$; $\Gamma^{H^0}_{tot}=0.14 \, GeV$ for
$m_{H^0}=300 \, GeV$; and $\Gamma^{A^0}_{tot}=0.045 \, GeV$ for
$m_{A^0}=300 \, GeV$ \cite{widths} and a luminosity
${\cal L}= 2 \times 10^{32} \, cm^{-2} \, s^{-1}$ \cite{emucoll}.
We present our numerical results for
$(\chi_{12})_{u.\,b.}^{e\mu\to\tau\tau}(s=m^2_{h^0})$ and
$(\chi_{12})_{u.\,b.}^{e\mu\to\tau\tau}(s=m^2_{H^0}=
m^2_{A^0})$ in Fig. 6 and Fig. 7, respectively.

We can also estimate the number of events
$N^{e\mu\to\tau\tau}(s)$
by taking for
$\chi_{12}$ the value for the current upper bound on
$\chi_{12}$ obtained from the measurements in $\mu^--e^-$
conversion experiments, namely
$\chi_{12}=(\chi_{12})_{u.\,b.}^
{\mu {\cal N} \to e {\cal N}}$ as a function of $\tan\beta$ for $Al$,
for the current experimental measurement $Br(\mu^- {\cal N} \to e^-
{\cal N}) < 6.1 \times 10^{-13}$ \cite{molzon}. Hence, we get 
\begin{equation}
N^{e\mu\to\tau\tau}(s)=
\sigma(e^- \mu^+ \to \tau^- \tau^+) \times \mbox{luminosity} \times 1
\, \mbox{year},
\end{equation}
as a function of $\alpha$ and $\tan\beta$.
In order to obtain numerical results, we take $\Gamma^{h^0}_{tot}=0.004
\, GeV$ for $m_{h^0}=115 \, GeV$; $\Gamma^{H^0}_{tot}=0.14 \, GeV$ for
$m_{H^0}=300 \, GeV$; and $\Gamma^{A^0}_{tot}=0.045 \, GeV$ for
$m_{A^0}=300 \, GeV$ \cite{widths} and a luminosity
${\cal L}= 2 \times 10^{32} \, cm^{-2} \, s^{-1}$ \cite{emucoll}.
We present our numerical results for
$N^{e\mu\to\tau\tau}(s=m^2_{h^0})$ and $N^{e\mu\to\tau\tau}(s=m^2_{H^0}=
m^2_{A^0})$ in Fig. 8 and Fig. 9, respectively. We obtain around  100
events {\it per} year, which is very likely detectable.

\section{Conclusions}

We have studied in this paper the lepton-Higgs couplings that
arise in the THDM-III, using a Hermitic four-texture form for the
leptonic Yukawa matrix. Because of this, although the fermion-Higgs
couplings are complex, the CP-properties of $h^0, H^0$ (even) and $A^0$
(odd) remmain valid.

We have derived bounds on the LFV parameters of the model, using current
experimental bounds on LFV transitions. Our resulting bounds can be
summarized as follows.
\begin{itemize}
\item $\chi_{12} < 5 \times 10^{-1}$, from $\mu^--e^-$ conversion
experiments.

\item $\chi_{13} = \chi_{23} < 6 \times 10^{-1}$,
from the radiative decay $\mu^+ \to e^+ \gamma$ measurements.
\end{itemize}

\noindent However, one can say that the present bounds on the
couplings $\chi_{ij}$'s still allow the possibility to study
interesting direct LFV Higgs signals at future colliders.

In particular, the LFV  couplings of the neutral Higgs bosons, can lead
to new discovery signatures of the Higgs boson itself. For instance,
the branching fraction for $H/A \to \tau\mu$ can be as large as $10^{-2}$,
while $Br(h \to \tau\mu)$ is also about
$10^{-2}$. These LFV Higgs modes complement the modes $B^0\to\mu\mu$,
$\tau \to 3\mu$, $\tau\to\mu\gamma$ and $\mu\to e\gamma$, as probes of
flavor violation in the THDM-III, which could provide key insights into
the form of the Yukawa mass matrix.

On the other hand, one can also relate our results with the SUSY-induced
THDM-III, by considering the effective Lagrangian for the couplings of
the charged leptons to the neutral Higgs fields, namely:
\beq
-{\cal L}=\bar{L}_L Y_l l_{R} \phi_1^0 +\bar{L}_L Y_l \left(\eps_1{\bf 1}
+\eps_2 Y_\nu^\dagger Y_\nu\right) l_{R} \phi_2^{0*} + h.c.
\label{FCLag}
\eeq
In this language, LFV results from our inability to simultaneously
diagonalize the term $Y_l$ and the non-holomorphic loop corrections, 
$\eps_2 Y_l Y_\nu^\dagger Y_\nu$. Thus, since the charged lepton masses
cannot be diagonalized in the same basis as their Higgs couplings, this
will allow neutral Higgs bosons to mediate LFV processes with rates
proportional to $\eps_2^2$. In terms of our previous notation we have:
$\tilde{Y}_2 = \eps_2 Y_l {Y_\nu}^\dagger {Y_\nu}$. An study of the values
for $\epsilon_2$ resulting from general soft breaking terms in the MSSM
is underway.

\bigskip

\noindent{\bf Acknowledgments.}

\noindent J. L. D.-C. and A. R. would like to thank {\it Sistema Nacional de
Investigadores} (M\'exico) for financial support, and the Huejotzingo
Seminar for inspiration. R. N.-P. acknowledges a graduate studies fellwoship
from CONACYT (M\'exico). This research was supported in part by CONACYT
(M\'exico).

\newpage

\begin{center}
{\bf Figure Captions}
\end{center}

\noindent{\bf Fig. 1}: The upper bound $(\chi_{23})_{u.\,b.}^{\tau \to 3\mu}$
as a function of $\tan\beta$ for $\alpha = \beta - \pi /4$,
$\alpha = \beta - \pi /3$, $\alpha = \beta - \pi /2$,
with $Br(\tau^- \to \mu^- \mu^+ \mu^-) < 1.9 \times 10^{-6}$,
taking  $m_{h^0} = 115$ $GeV$ and $m_{H^0} = m_{A^0} = 300$ $GeV$.

\bigskip

\noindent{\bf Fig. 2}: The upper bound $(\chi_{23})_{u.\,b.}
^{\mu \to e \gamma}$
as a function of $\tan\beta$ for $\alpha = \beta - \pi /4$,
$\alpha = \beta - \pi /3$, $\alpha = \beta - \pi /2$,
with $Br(\mu^+ \to e^+ \, \gamma) < 1.2 \times 10^{-11}$,
taking  $m_{h^0} = 115$ $GeV$ and $m_{H^0} = m_{A^0} = 300$ $GeV$.

\bigskip

\noindent{\bf Fig. 3}: The upper bound $(\chi_{12})_{u.\,b.}^
{\mu {\cal N} \to e {\cal N}}$ as a function of $\tan\beta$ for $Al$,
$Pb$, with $Br(\mu^- {\cal N} \to e^- {\cal N}) < 6.1 \times 10^{-13}$ and
assuming $m_{H^0} = 300$ $GeV$.

\bigskip

\noindent{\bf Fig. 4}: The same as in Fig. 3, but taking
$Br(\mu^- {\cal N} \to e^- {\cal N}) < 2 \times 10^{-17}$.

\bigskip

\noindent{\bf Fig. 5}: $\Delta a_{\mu}^{h^0}$ as a function of
$\tan\beta$ for $\alpha = \beta - \pi /4$, $\alpha = \beta =
- \pi /3$, with $\chi_{23}=1$ and assuming $m_{h^0} = 115$ $GeV$.

\bigskip

\noindent{\bf Fig. 6}: The upper bound $(\chi_{12})_{u.\,b.}^
{e\mu\to\tau\tau}$ for $s=m^2_{h^0}=(115 \, GeV)^2$,
with $\Gamma^{h^0}_{tot}=0.004 \, GeV$, as a function of $\tan\beta$ for
$\alpha = \beta - \pi /4$, $\alpha = \beta - \pi /3$,
when $\sigma(e^- \mu^+ \to \tau^- \tau^+)
\times \mbox{luminosity} \times 1 \, \mbox{year} < 1$, taking
${\cal L}= 2 \times 10^{32} \, cm^{-2} \, s^{-1}$

\bigskip

\noindent{\bf Fig. 7}: The upper bound $(\chi_{12})_{u.\,b.}^
{e\mu\to\tau\tau}$ for $s=m^2_{H^0}=m^2_{A^0}=(300 \, GeV)^2$,
with $\Gamma^{H^0}_{tot}=0.14 \, GeV$
and $\Gamma^{A^0}_{tot}=0.045 \, GeV$,
as a function of $\tan\beta$ for
$\alpha = \beta - \pi /4$, $\alpha = \beta - \pi /3$,
$\alpha = \beta - \pi /2$, when $\sigma(e^- \mu^+ \to \tau^- \tau^+)
\times \mbox{luminosity} \times 1 \, \mbox{year} < 1$, taking
${\cal L}= 2 \times 10^{32} \, cm^{-2} \, s^{-1}$

\bigskip

\noindent{\bf Fig. 8}: Number of events $N^{e\mu\to\tau\tau}$
for $s=m^2_{h^0}=(115 \, GeV)^2$,
taking $\chi_{12} =(\chi_{12})_{u.\,b.}^ {\mu Al \to e Al}$
for $Br(\mu^- Al \to e^- Al) < 6.1 \times 10^{-13}$
with $\Gamma^{h^0}_{tot}=0.004 \, GeV$, as a function of $\tan\beta$ for
$\alpha = \beta - \pi /4$, $\alpha = \beta - \pi /3$,
taking ${\cal L}= 2 \times 10^{32} \, cm^{-2} \, s^{-1}$

\bigskip

\noindent{\bf Fig. 9}: Number of events $N^{e\mu\to\tau\tau}$
for $s=m^2_{H^0}=m^2_{A^0}=(300 \, GeV)^2$,
taking $\chi_{12} =(\chi_{12})_{u.\,b.}^ {\mu Al \to e Al}$
for $Br(\mu^- Al \to e^- Al) < 6.1 \times 10^{-13}$
with $\Gamma^{H^0}_{tot}=0.14 \, GeV$
and $\Gamma^{A^0}_{tot}=0.045 \, GeV$,
as a function of $\tan\beta$ for
$\alpha = \beta - \pi /4$, $\alpha = \beta - \pi /3$,
$\alpha = \beta - \pi /2$, taking
${\cal L}= 2 \times 10^{32} \, cm^{-2} \, s^{-1}$


\begin{thebibliography}{99}

\bibitem{hixradc} For a review see:
U.~Baur {\it et al.}  [The Snowmass Working Group on Precision Electroweak
                  Measurements Collaboration],
in {\it Proc. of the APS/DPF/DPB Summer Study on the Future of Particle 
Physics (Snowmass 2001) } ed. N.~Graf,
eConf {\bf C010630}, P1WG1 (2001)
[arXiv:hep-ph/0202001].

\bibitem{hixphen} For a review see:
S. Dawson {\it et al.}, ``The Higgs Hunter's Guide''.

\bibitem{susyrev} 
See, for instance, recent reviews in ``Perspectives on Supersymmetry'',
ed. G.\,L. Kane, World Scientific Publishing Co., 1998;
H.\,E. Haber, Nucl. Phys. Proc. Suppl. {\bf 101}, 217 (2001),
[hep-ph/0103095].  

\bibitem{deconst} N. Arkani-Hamed, A. Cohen and H. Georgi,
Phys. Lett. B{\bf 513} (2001) 232 [arXiv: hep-h/0105239].

\bibitem{mssmhix} For a review of MSSM Higgs phenomenology see:
M.~Carena {\it et al.},
``Report of the Tevatron Higgs working group,''
arXiv:hep-ph/0010338;
see also: C.~Balazs, J.~L.~Diaz-Cruz, H.~J.~He, T.~Tait and C.~P.~Yuan,
Phys.\ Rev.\ D {\bf 59}, 055016 (1999)
[arXiv:hep-ph/9807349];
J.~L.~Diaz-Cruz, H.~J.~He, T.~Tait and C.~P.~Yuan,
Phys.\ Rev.\ Lett.\  {\bf 80}, 4641 (1998)
[arXiv:hep-ph/9802294].

\bibitem{seesaw}
M.~Gell-Mann, P.~Ramond and R.~Slansky, in {\it Supergravity},
eds.~P.~van Nieuwenhuizen and D.Z.~Freedman (North Holland 1979);
T.~Yanagida, in Proceedings of the {\it Workshop on Unified Theory and
Baryon Number in the Universe}, eds.~O.~Sawada and A.~Sugamoto (KEK
1979); 
R.N.~Mohapatra and G.~Senjanovi\'c, Phys.\ Rev.\ Lett.\ {\bf 44}, 912
(1980). 

\bibitem{atmospheric}
S.~Fukuda \etal, (SuperKamiokande Collaboration), Phys.\ Rev.\ Lett.\
{\bf 85}, 3999 (2000).

\bibitem{solar}
S.~Fukuda \etal, (SuperKamiokande Collaboration), Phys.\ Rev.\ Lett.\
{\bf 86}, 5656 (2001); Q.R.~Ahmad \etal, (SNO Collaboration), 
Phys.\ Rev.\ Lett.\ {\bf 87}, 071301 (2001) and 
Phys.\ Rev.\ Lett.\ {\bf 89}, 011301 (2002).

\bibitem{bakotau} K.~S.~Babu and C.~Kolda,
Phys.\ Rev.\ Lett.\  {\bf 89}, 241802 (2002)
[arXiv:hep-ph/0206310];
A.~Dedes, J.~R.~Ellis and M.~Raidal,
Phys.\  Lett.\  {\bf B549}, 159 (2002)
[arXiv:hep-ph/0209207].

\bibitem{myhlfvA} J.~L.~Diaz-Cruz and J.~J.~Toscano,
Phys.\ Rev.\ D {\bf 62}, 116005 (2000)
[arXiv:hep-ph/9910233];
J.~L.~Diaz-Cruz,
JHEP {\bf 0305}, 036 (2003)
[arXiv:hep-ph/0207030].

\bibitem{prlbako}
K.~S.~Babu and C.~Kolda,
Phys.\ Rev.\ Lett.\  {\bf 84}, 228 (2000).

\bibitem{bmuanalysis}
C.~S.~Huang, W.~Liao, Q.~S.~Yan and S.~H.~Zhu,
Phys.\ Rev.\ D {\bf 63}, 114021 (2001);
A.~Dedes, H.~K.~Dreiner and U.~Nierste,
Phys.\ Rev.\ Lett.\  {\bf 87}, 251804 (2001);
G.~Isidori and A.~Retico,
JHEP {\bf 0111}, 001 (2001);
R.~Arnowitt, B.~Dutta, T.~Kamon and M.~Tanaka,
Phys.\ Lett.\ B {\bf 538}, 121 (2002);
C.~Bobeth \etal, arXiv:hep-ph/0204225;
H.~Baer \etal, arXiv:hep-ph/0205325.

\bibitem{FN}
C. D. Frogatt and H. B. Nielsen,
Nucl. Phys. B{\bf 147}, 277 (1979).

\bibitem{fritzsch} H. Fritzsch, Phys. Lett. B70 (1977) 436.

\bibitem{chengsher} T.P. Cheng and M. Sher,
Phys.\ Rev.\ D {\bf 35} (1987) 3484;

\bibitem{others} A.~Antaramian, L.~J.~Hall and A.~Rasin,
Phys.\ Rev.\ Lett.\  {\bf 69}, 1871 (1992);
M.~Sher and Y.~Yuan,
Phys.\ Rev.\ D {\bf 44}, 1461 (1991).

\bibitem{muchos}
J.L. Diaz-Cruz {\it et al.}, Phys. Rev. {\bf D51} 5263 (1995);
J.L. Diaz-Cruz and G. Lopez Castro, Phys. Lett. {\bf B301}, 405 (1993);
M. Luke and M. Savage, Phys. Lett. {\bf B307}, 387 (1993);
D. Atwood {\it et al.}, Phys. Rev. {\bf D55},3156 (1997).


\bibitem{fourtext} H. Fritzsch and Z. Z. Xing, Phys. Lett. {\bf 555}, 63
(2003) (arXiv: hep-ph/0212195).

\bibitem{cpv}
K.~S.~Babu, C.~Kolda, J.~March-Russell and F.~Wilczek,
Phys.\ Rev.\ D {\bf 59}, 016004 (1999).

\bibitem{zhou} Y-F. Zhou, arXiv: hep-ph/0307240.

\bibitem{exp}
F.~Deppisch {\it et al.}, Eur. Phys. J. {\bf C28}, 365 (2003)
[arXiv: hep-ph/0206122] and references therein.

\bibitem{chang1} D. Chang, W. S. Hou  and W.-Y. Keung, Phys. Rev. {\bf D48},
217 (1993).

\bibitem{partdata} K. Hagiwara {\it et al.} (Particle Data Group), Phys. Rev.
{\bf D66}, 010001 (2002).

\bibitem{psiexp} {\it MUEGAMMA Collaboration}, Nucl. Instrum. Meth.
{\bf A503}, 287 (2003).

\bibitem{kitano} R. Kitano {\it et al.}, Phys. Lett. {\bf B575}, 300
(2003) [arXiv:hep-ph/0308021].

\bibitem{suzuki} T. Suzuki, D. F. Measday and J. P. Roalsvig, Phys. Rev.
{\bf C35}, 2212 (1987).

\bibitem{barkov} L. M. Barkov {\it et al.}, research proposal to PSI; S.Ritt,
in Proc. of {\it The 2nd International Workshop on Neutrino Oscillations and
their Oigin}, edited by Y. Suzuki {\it et al.} (World Scientific), p. 245
(2000).

\bibitem{bachman} M. Bachman {\it et al.} [MECO Collaboration], experimental
proposal E940 to Brookhaven National Laboratory AGS (1997).

\bibitem{kuno} Y. Kuno, in Proc. of {\it The 2nd International Workshop on
Neutrino Oscillations and thei Oigin}, edited by Y. Suzuki {\it et al.}
(World Scientific), p. 245 (2000).

\bibitem{kosmas} T. S. Kosmas in Proc. of {\it Joint U.S./Japan Workshop
on New Initiatives in Muon Lepton Flavor Violation and Neutrino Oscillation
with High Intense Muon and Neutrino Sources}, Honolulu, Hawaii (2000). p. 64.

\bibitem{molzon} W. Molzon, Nucl. Phys. {\bf B111} - Proc. Suppl., 188 (2002).

\bibitem{zhou-wu} Y.-F. Zhou and Y.-L. Wu, Eur. Phys. J. {\bf C27}, 577 (2003).

\bibitem{bennett} G. W. Bennett, Phys. Rev. Lett. {\bf 89}, 101804 (2002)
[arXiv: hep-ex/0208001].

\bibitem{amutheo1} F. Jegerlehner, talk given at the Workshop Centre de
Physique Theorique Marseille, France, 14-16 March 2002; F. Jegerlehner,
[arXiv: hep-ph/0310234].

\bibitem{amutheo2} K. Hagiwara {\it et al.}, talk given at ICHEP'02,
Amsterdam, Netherlands, 24-31 July 2002. Published in {\it Amsterdam 2002,
ICHEP}, p. 216; K. Hagiwara {\it et al.}, [arXiv: hep-ph/0312250].

\bibitem{amu1} S. Nie and M. Sher, Phys. Rev. {\bf D58}, 097701 (1998).

\bibitem{amu2} S. K. Kang and K. Y. Lee, Phys. Lett. {\bf B521}, 61 (2001),
[arXiv: hep-ph/0103064].

\bibitem{amu3} R. Diaz, R. Martinez and J. A. Rodriguez,
Phys. Rev. {\bf D63}, 095007 (2001) [arXiv: hep-ph/0010149].

\bibitem{emucoll} V. Barger, S. Pakvasa and X. Tata, Phys. Lett. {\bf B415},
200 (1997).

\bibitem{myhlfvB}
U.~Cotti, L.~Diaz-Cruz, C.~Pagliarone and E.~Vataga,
in {\it Proc. of the APS/DPF/DPB Summer Study on the Future of Particle Physics (Snowmass 2001) } ed. N.~Graf,
eConf {\bf C010630}, P102 (2001)
[arXiv:hep-ph/0111236];
U.~Cotti, J.~L.~Diaz-Cruz, R.~Gaitan, H.~Gonzales and A.~Hernandez-Galeana,
Phys.\ Rev.\ D {\bf 66}, 015004 (2002)
[arXiv:hep-ph/0205170].

\bibitem{otherlfvh} 
M.~Sher,
Phys.\ Lett.\ B {\bf 487}, 151 (2000)
[arXiv: hep-ph/0006159];
A.~Brignole and A.~Rossi, Phys. Lett. {\bf 566}, 217 (2003)
[arXiv: hep-ph/0304081].

\bibitem{hspira} M. Spira, Nucl. Instrum. Meth. {\bf A389}, 357 (1997)
[arXiv: hep-ph/910350].

\bibitem{htaumubkd}
T.~Han and D.~Marfatia,
Phys.\ Rev.\ Lett.\  {\bf 86}, 1442 (2001)
[arXiv:hep-ph/0008141];
K.~A.~Assamagan, A.~Deandrea and P.~A.~Delsart,
Phys.\ Rev.\ D {\bf 67}, 035001 (2003)
[arXiv:hep-ph/0207302].

\bibitem{widths} A. Djouadi, J. Kalinowski and M. Spira,
Comput. Phys. Commun. {\bf 108}, 56 (1998) [arXiv: hep-ph/9704448].

\end{thebibliography}
\end{document}